\def\o{\over}
\def\A{\rightarrow}
\def\bar{\overline}

\def\r{\gamma}
\def\a{\alpha}
\def\b{\beta}
\def\e{\epsilon}
\def\p{\pi}
\def\Im{\rm Im}

\def\t{\tilde}
\def\bar{\overline}

\def\G{{\rm GeV}}
\documentstyle [12pt]{article}
\begin{document}
\baselineskip=25pt
\setcounter{page}{1}
\thispagestyle{empty}
\topskip 0.2  cm
\begin{flushright}
\begin{tabular}{c c}
& {\normalsize   AUE-02-94}\\
& {\normalsize  EHU-01-94}\\
& {\normalsize  US-94-01}\\
& {\normalsize  January 1994}
\end{tabular}
\end{flushright}
\vspace{-0.2cm}
\centerline{\Large\bf Neutron Electric Dipole Moment}
 \centerline{\Large\bf in  Two Higgs Doublet Model}
\vskip 0.3 cm
\centerline{{\bf Takemi HAYASHI}$^{(a)}$, \hskip 1 cm
           {{\bf Yoshio KOIDE}$^{(b)}$
  \footnote{E-mail:koide@u-shizuoka-ken.ac.jp} }}
\vskip -0.3cm
\centerline{and}
\vskip -0.3cm
\centerline{{{\bf Masahisa  MATSUDA}$^{(c)}$,
              \footnote{E-mail:masa@auephyas.aichi-edu.ac.jp}} \hskip 1 cm
           {\bf Morimitsu TANIMOTO}$^{(d)}$}
 \vskip 0.1 cm
\centerline{$^{(a)}$ \it{Kogakkan University, Ise, Mie 516, JAPAN}}
\centerline{$^{(b)}$ \it{Department of Physics, University of Shizuoka,
  52-1 Yada, Shizuoka 422, JAPAN}}
\centerline{$^{(c)}$ \it{Department of Physics and Astronomy, Aichi University
of Education}}
\centerline{\it Kariya, Aichi 448, JAPAN}
\centerline{$^{(d)}$ \it{Science Education Laboratory, Ehime University,
 Matsuyama 790, JAPAN}}
\vskip 0.3 cm
\centerline{\bf ABSTRACT}
We study  the effect of the "chromo-electric" dipole moment  on
the electric dipole moment(EDM) of the neutron in the two Higgs doublet model.
   We systematically investigate the Weinberg's operator $O_{3g}=GG\t G$
and the operator $O_{qg}=\bar q\sigma\t Gq$,
in the cases of $\tan\b\gg 1$, $\tan\b\ll 1$ and $\tan\b\simeq 1$.
  It is shown that $O_{sg}$  gives the main contribution to
  the neutron EDM compared to the  other operators,
 and also that
the contributions of $O_{ug}$ and $O_{3g}$ cancel out each other.
It is pointed out that the inclusion of second lightest neutral Higgs  scalar
adding to the lightest one is of essential importance to estimate
  the neutron  EDM.
  The neutron EDM is considerably reduced due to the destructive
contribution with each other
if the mass difference of the two Higgs scalars is of the order  $O(50\G)$.
\par
\newpage
\topskip 1 cm
\noindent
{\bf 1. Introduction}\par

  The physics of  $CP$ violation has attracted one's attention
in the circumstance that the $B$-factory will go on in the near future.
In the experiments of the $B$ decay asymmetry, the central subject is
the test of the standard Kobayashi-Maskawa model(SM)[1] as an origin of $CP$
violation.
  On the other hand,
 the electric dipole moment(EDM) of the neutron is of central importance
to probe a new origin  of $CP$ violation, because it is very small in SM.
Begining with the papers of Weinberg[2], there has been considerable renewed
interest in the neutron EDM induced by $CP$ nonconservation of the neutral
   Higgs sector.
Some studies[3,4,5] revealed numerically the importance of the
"chromo-electric" dipole moment, which arises from the three-gluon operator
$GG\t G$ by found Weinberg[2] and the light quark operator
$\bar q \sigma\t Gq$  introduced
by Gunion and Wyler[3], in the neutral Higgs sector.
Thus, it is important to study the effect of these operators systematically
in the model beyond  SM.
In this paper, we study the contribution of above two operators to
the neutron EDM in the two Higgs doublets model(THDM)[6].
  The $3\times 3$ mass matrix of the neutral Higgs scalars is
carefully investigated in the typical three cases of $\tan\b\gg 1$,
$\tan\b\simeq 1$ and $\tan\b\ll 1$, where $\tan\b\equiv |v_2/v_1|$ with
$v_i\equiv \langle \phi_i^0\rangle_{vac}$ and $\phi_1$ and $\phi_2$ couple
with down- and up-quark sectors, respectively.
 In these restricted regions of $\tan\b$,
       the Higgs mass matrix becomes very simple,    and then
we can easily estimate the $CP$ violation parameters of
the neutral Higgs sector, which lead to the neutron EDM in THDM.
We found that the neutron EDM  follows mainly from the two light
neutral Higgs  scalar exchanges.  Due to the opposite signs of the
two contributions,  the neutron EDM is considerably reduced
 if the mass difference of the two Higgs scalars is in the order of $O(50\G)$.
\par
  In order to give reliable predictions, one needs the improvement on the
accuracy of the description of the strong-interaction hadronic  effects.
Weinberg employed
the "naive dimensional analyse"(NDA) as developed by Georgi and Manohar[7]
 in computing the effect of the $GG\t G$ operator on the neutron.
However, this method admittedly provides at best the order-of-magnitude
estimation.
Moreover, when gluon fields are present, there occurs an indeterminable factor
of $4\p$, which depends on whether one associates a factor $g_s$ or $4\p g_s$
with each gluon field factor in the interaction Lagrangian[3].
Recently, Chemtob[8] proposed a systematic approach
which gives the hadronic matrix elements of the higher-dimension operators
involving the gluon fields by using the large $N_c$ current-algebra.
In his model,
 the hadronic matrix elements of the operators
 are approximated by the intermediate states with the single nucleon pole
 and the  nucleon plus one pion. So, this approach may be a realistic one.
We employ his model to get the hadronic matrix elements of the relevant
operators  in this work.
The comparision between results by this approach and NDA will be discussed
 briefly in the last section.\par
In section 2, the neutral Higgs mass matrix is analyzed and then
the magnitudes of the $CP$ violation factors ${\rm Im} Z_i$ are estimated.
In section 3, the formulation of the neutron EDM with the hadronic
matrix elements of the $CP$ violating  operators are discussed.
Section 4 is devoted to the numerical results of the neutron EDM
and  some remarks and conclusion are given in section 5.\par
\vskip 0.5 cm
\noindent
{\bf 2. $CP$ violation parameter in  THDM}\par

The simple extension of SM is the one with the two Higgs doublets[6].
This model      has the possibility
of the soft $CP$ violation in the neutral Higgs sector, which
does not contribute to the flavor changing neutral current in the $B$, $D$
and $K$ meson decays.  Weinberg[9] has given the unitarity bounds for
the dimensionless parameters of the $CP$ nonconservation in THDM.
However, values of these parameters are not always  close to
  the Weinberg's bounds[9]. Actually,  the $CP$ violation
parameter ${\rm Im}Z_1$(this definition is given later)
 is suppressed by $1/\tan\b$ compared with the Weinberg's bound
at the large $\tan\b$ as pointed out by Barr[5].
 Chemtob[10]  has predicted $CP$ violation parameters
 by the use of the renormalization approach
under the assumption that the coupling constants of the Yukawa couplings
and self-coupling scalar mesons interactions reach infrared fixed points at the
 electroweak scale.
 This infrared fixed points approach either leads to the
 large top quark mass $m_t\sim 230\G$, which is unfavourable to the recent
electroweak precision test, or leads to the existence of
the unobserved fourth generation quarks. Thus, it is difficult to estimate the
reliable magnitudes of the $CP$ violation parameters ${\rm Im} Z_i(i=1,2)$.
However, we found that the Higgs mass matrix is simplified in the extreme
cases of  $\tan\b\ll 1$, $\tan\b\simeq 1$ and $\tan\b\gg 1$, in which the
$CP$ violation parameters are easily calculated.\par
The $CP$ violation will occur via the scalar-pseudoscalar interference terms
involving the imaginary parts of the scalar meson fields normalization
constants, $Z_i$, which are column vectors in the neutral Higgs scalar vector
space, defined in terms of the tree level approximation
to the two-point function as follows:
\begin{eqnarray}
& &\left[\matrix{{1\o v_1^2}\langle \phi_1^0\phi_1^0\rangle_q,&
{1\o v_2^2}\langle \phi_2^0\phi_2^0\rangle_q,&
{1\o v_1v_2}\langle \phi_2^0\phi_1^0\rangle_q,&
{1\o v_1^*v_2}\langle \phi_2^{0}\phi_1^{*0}\rangle_q}\right]  \nonumber \\
       & &=\sum_{n=1}^3 {\sqrt{2}G_F\o q^2-m^2_{Hn}}
          [\matrix{Z_1^{(n)},&Z_2^{(n)},&\tilde Z_0^{(n)},&Z_0^{(n)}}] \ ,
\end{eqnarray}
\noindent  where $v_i\equiv \langle \phi_i^0\rangle_{vac}$.
The $CP$ violation factors ${\rm Im} Z_i^{(n)}$ are deduced to
\begin{eqnarray}
  {\rm Im} Z_2^{(k)}&=&{1\o \tan\b\sin\b}u_2^{(k)} u_3^{(k)} \ , \quad
  {\rm Im} Z_1^{(k)}=-{\tan\b\o \cos\b}u_1^{(k)} u_3^{(k)} \ , \\
  {\rm Im} \tilde Z_0^{(k)}&=&{1 \o 2}\left ({1\o \sin\b}u_1^{(k)} -
                {1\o \cos\b} u_2^{(k)}\right ) u_3^{(k)}\ , \quad
  {\rm Im} Z_0^{(k)}={1 \o 2}\left ({1\o \sin\b}u_1^{(k)} +
                {1\o \cos\b} u_2^{(k)}\right ) u_3^{(k)}\ , \nonumber
\end{eqnarray}
\noindent
where $u_i^{(k)}$ denotes the $i-$th component of the $k-$th normalized
eigenvector of the Higgs mass matrix.
 Let us estimate $u_i^{(k)}$ by studying the symmetric
Higgs mass matrix ${\bf M^2}$
whose components are
\begin{eqnarray}
M_{11}^2&=&2g_1|v_1|^2+g'|v_2|^2+{\xi+Re(hv_1^{*2}v_2^2)\o|v_1|^2}\ ,
        \nonumber\\
M_{22}^2&=&2g_2|v_2|^2+g'|v_1|^2+{\xi+Re(hv_1^{*2}v_2^2)\o|v_2|^2}\ ,
        \nonumber\\
M_{33}^2&=&(|v_1|^2+|v_2|^2) \left [g'+
              {\xi-Re(hv_1^{*2}v_2^2)\o|v_1v_2|^2}\right ]\ ,\nonumber\\
M_{12}^2&=&|v_1v_2|(2g+g')+{Re(hv_1^{*2}v_2^2)-\xi\o|v_1v_2|}\ , \\
M_{13}^2&=&-{\sqrt{|v_1|^2+|v_2|^2}\o|v_1^2v_2|}\Im(hv_1^{*2}v_2^2)\ ,
         \nonumber\\
M_{23}^2&=&-{\sqrt{|v_1|^2+|v_2|^2}\o|v_1v_2^2|}\Im(hv_1^{*2}v_2^2)\ ,\nonumber
\end{eqnarray}
which are derived from the Higgs potential
\begin{eqnarray}
V&=&{1\o 2}g_1(\phi_1^\dagger\phi_1-|v_1|^2)^2+
         {1\o 2}g_2(\phi_2^\dagger\phi_2-|v_2|^2)^2 \nonumber\\
  &+& g(\phi_1^\dagger\phi_1-|v_1|^2)(\phi_2^\dagger\phi_2-|v_2|^2)
      \nonumber\\
&+& g'|\phi_1^\dagger\phi_2-v_1^*v_2|^2+Re[h(\phi_1^\dagger\phi_2-v_1^*v_2)^2]
        \nonumber\\
  &+& \xi\left [{\phi_1\o v_1}-{\phi_2\o v_2}\right ]^\dagger
         \left [{\phi_1\o v_1}-{\phi_2\o v_2}\right ] \ .
\end{eqnarray}
\noindent
As a phase convension, we take $h$ to be real and
\begin{equation}
  v_1^{*2} v_2^2=|v_1|^2|v_2|^2\exp(2i\phi) \ .
\end{equation}
Now, the Higgs mass matrix ${\bf M^2}$ is rotated so as to make
 the (1,3)(and  then (3,1))
component  zero  by the orthogonal matrix ${\bf U_0}$ as
\begin{equation}
 {\bf U_0}=\left(\matrix{\cos\b& \sin\b &0\cr -\sin\b& \cos\b &0\cr 0&0&1 }
   \right)\ .
\end{equation}
\noindent
Then, the transformed matrix ${\bf M'^2=U_0^t M^2 U_0}$ is given as
\begin{eqnarray}
M_{11}^{'2}&=&2g_1\cos^4\b+2g_2\sin^4\b+4(\bar\xi-g)\sin^2\b\cos^2\b\
,\nonumber\\
M_{22}^{'2}&=&2(g_1+g_2+2g-2\bar\xi)\sin^2\b\cos^2\b\
        +g'+\bar\xi+h\cos2\phi \ ,         \nonumber\\
M_{33}^{'2}&=& g'+ \bar\xi-h\cos 2\phi \ ,\nonumber\\
M_{12}^{'2}&=&\sin\b\cos\b \left[\cos 2\b(2g-2\bar\xi+g_1+g_2)+g_1-g_2
          \right ]\ ,  \\
M_{13}^{'2}&=&0 \ ,         \nonumber\\
M_{23}^{'2}&=&-h\sin2\phi \ ,\nonumber
\end{eqnarray}
\noindent
in the $v^2\equiv |v_1|^2+|v_2|^2$ unit and the parameter $\bar\xi$ is
                defined as   $\bar\xi=\xi/|v_1v_2|^2$.
This matrix cannot be diagonalized in the analytic form generally, unless
 special relations among the parameters of the mass matrix are satisfied.
 The parameters are only constrained by the positivity condition as
follows[10,11]:
\begin{equation}
g_1>0\ ,\quad g_2>0\ ,\quad h<0\ ,\quad h+g'<0\ ,\quad g+g'+h>-\sqrt{g_1g_2}\ .
\end{equation}
However, we can simply diagonalize the Higgs mass matrix
 in the extreme cases  of
 $\tan\b\gg 1$, $\tan\b\simeq 1$ and $\tan\b\ll 1$.\par
At first, we consider the case of $\tan\b\gg 1$.
By retaining the order of $\cos\b$ and by setting $\cos^2\b=0$, $\sin\b=1$,
the mass matrix becomes
\begin{equation}
 \left ( \matrix{2g_2 & 2\cos\b(\bar\xi-g-g_2) &0 \cr
       2\cos\b(\bar\xi-g-g_2) & g'+\bar\xi+h\cos2\phi & -h\sin2\phi \cr
         0 & -h\sin2\phi & g'+\bar\xi-h\cos2\phi} \right ) \ .
\end{equation}
\noindent
In the limit of $\cos\b=0$,  this mass matrix is diagonalized by only rotating
 $\phi$ on the (2-3) plane. However,
 due to the non-vanishing tiny $M_{12}^{'2}$ component,
this rotation is slightly deviated from the (2-3) plane.
The orthogonal matrix ${\bf U_1}$ to diagonalize the Higgs mass matrix of
Eq.(9)  is approximately obtained as:
\begin{equation}
 {\bf U_1}\simeq\left ( \matrix{1 & 0 &0 \cr 0 & \cos\phi & \sin\phi \cr
         0 & -\sin\phi & \cos\phi} \right )
 \left(\matrix{1 & \e\cos\phi &0 \cr -\e\cos\phi & 1& 0 \cr 0 & 0& 1} \right )
 \left( \matrix{1 & 0 &\e\sin\phi\cr 0 & 1& 0 \cr -\e\sin\phi& 0& 1}\right )\ ,
\end{equation}
\noindent where, neglecting $h\cos 2\phi$,
\begin{equation}
 \e \simeq {2(\bar\xi-g-g_2)\o \bar\xi+g'-2g_2}\cos\b \ .
\end{equation}
Then, the eigenvectors of ${\bf M^2}$ in Eq.(1) are
\begin{eqnarray}
u^{(1)}&=&\{\matrix{\cos\b-\e\sin\b, &
             -\sin\b, & 0 }\} \ ,\nonumber\\
u^{(2)}&=&\{\matrix{\sin\b\cos\phi, & (\cos\b-\e\sin\b)\cos\phi,
              & -\sin\phi }\} \ , \\
u^{(3)}&=&\{\matrix{\sin\b\sin\phi, &(\cos\b -\e\sin\b)\sin\phi,
              & \cos\phi }\} \ , \nonumber
\end{eqnarray}
\noindent with the order of $O(\cos^2\b)$ being neglected.
The diagonal masses are given as
\begin{equation}
M_1^2=2g_2+O(\cos^2\b),\quad M_2^2=g'+\bar\xi+h+O(\cos^2\b),
       \quad M_3^2=g'+\bar\xi-h+O(\cos^2\b) \ ,
\end{equation}
\noindent               in the $v^2$ unit.
The lightest Higgs scalar to yield  $CP$ violation is the  second Higgs
  scalar with the mass $M_2$ since  $h$ is negative and
  $\bar\xi$ is positive.
 The Higgs scalar with $M_1$ does not contribute to  $CP$ violation
because of $u_3^{(1)}=0$. The absolute values of  $g'$
is expected to be $O(1)$, but $h$ seems to be small as estimated
in  some works[11].
 Therefore, the masses $M_2$ and $M_3$ may be almost degenerated. Then,
$CP$ violation is reduced by the cancellation between the two different
  Higgs  exchange contributions $\Im Z_i^{(2)}$ and $\Im Z_i^{(3)}$
since $u^{(2)}_i u^{(2)}_3$ and
  $u^{(3)}_i u^{(3)}_3$(i=1,2) have same magnitudes with opposite signs.
 Thus, it is noted that
      the lightest single Higgs exchange approximation gives miss-leading
of $CP$ violation in the case of $\tan\b\gg 1$.\par
In order to get the magnitudes of $u_2^{(2,3)}$,  we estimate $\e$, which
depends on the value of $\bar\xi$.
The parameter $\bar\xi$ is determined by the charged Higgs mass as follows:
\begin{equation}
M^{\pm 2}=\bar\xi v^2 \ .
\end{equation}
 We have already studied the charged Higgs scalar effect in THDM through
 the  inclusive decay $B\A X_s\r$[12], as to which the  upper bound of the
branching ratio was recently  given by the CLEO collaboration[13].
We obtained $300\G$ for the lower bound of the charged Higgs scalar mass
 in the case of $m_t=150\G$ and $m_b=5\G$. This lower bound means
 $\bar\xi>3$.
 In the limit of the large $\bar\xi$ with retaining
  other parameters to be $O(1)$,
 $\e/\cos\b$ reachs 2 as seen in Eq.(11). Actually,
$g_1$, $g_2$, $g$ and $|g'|$ are around 1 in some numerical studies[11].
Then, if we take $M^{\pm}=400(350)\G$,
which corresponds to $\bar\xi=5(4)$, the value of $\e/\cos\b$ becomes 3(4).
 In the following calculations, we fix to be $\e=3\cos \b$.
 By use of  these resulting $u^{(2)}_i$ values,
               we can calculate the $CP$ violation
factors ${{\Im}Z_i}={{\Im}Z_i^{(2)}}$. We show the numerical results
together with the Weinberg bounds
 for ${\Im}Z_1$ and ${\Im}Z_2$ in Figs.1(a) and 1(b),
 where $\phi=\p/4$ is taken.
 Although the Weinberg bounds give nothing for these signs,
 our estimates determine the relative sign between ${\Im}Z_1$ and ${\Im}Z_2$.
For  ${\Im}Z_1$, our result reachs the Weinberg bound, but for  ${\Im}Z_2$
 the our calculated value is suppressed compared with
 the Weinberg bound in the order of  $1/\tan\b$.
\begin{center}
 \unitlength=0.7 cm
 \begin{picture}(2.5,2.5)
  \thicklines
  \put(-2,0){\framebox(6,1){\bf Figs.1(a) $\sim$ 1(f)}}
 \end{picture}
\end{center}
\par
$CP$ violation in the case of $\tan\b\ll 1$ is similar to the one of
              $\tan\b\gg 1$.
By retaining the order of $\sin\b$ and by setting $\sin^2\b=0$, $\cos\b=1$,
the mass matrix becomes
\begin{equation}
 \left ( \matrix{2g_1 & -2\sin\b(\bar\xi-g-g_1) &0 \cr
         -2\sin\b(\bar\xi-g-g_1) & g'+\bar\xi+h\cos2\phi & -h\sin2\phi \cr
         0 & -h\sin2\phi & g'+\bar\xi-h\cos2\phi} \right ) \ .
\end{equation}
Except for replacing $g_2$ with $g_1$ and  $\cos\b$ with $-\sin\b$,
the mass matrix is the same one as of Eq.(9) in the case of $\tan\b\ll 1$.
The eigenvectors are easily obtained as follows:
\begin{eqnarray}
u^{(1)}&=&\{\matrix{\cos\b, &
             -(\sin\b+\e'\cos\b), & 0 }\} \ ,\nonumber\\
u^{(2)}&=&\{\matrix{(\e'+\sin\b)\cos\phi, & \cos\b\cos\phi,
              & -\sin\phi }\} \ , \\
u^{(3)}&=&\{\matrix{(\e'+\sin\b)\cos\phi, & \cos\b \sin\phi,
              & \cos\phi }\} \ , \nonumber
\end{eqnarray}
\noindent with the order of $O(\sin^2\b)$ being neglected. The $\e'$ parameter
is defines as
\begin{equation}
 \e'=-{2(\bar\xi-g-g_1)\o \bar\xi+g'-2g_1+h\cos2\phi}\sin\b \ ,
\end{equation}
\noindent
which is taken to be $\e'=-3\sin\b$ as discussed in Eq.(11).
 Taking $u^{(2)}_i$ as the eigenvector of the lightest Higgs scalar,
 we show  ${\Im}Z_1$ and ${\Im}Z_2$ in Figs.1(c) and 1(d).
For  ${\Im}Z_2$, our result reachs the Weinberg bound, while for  ${\Im}Z_1$
 the calculated value is suppressed from the Weinberg bound in the order of
 $\tan\b$. The relative sign between ${\Im}Z_1$ and ${\Im}Z_2$ is just
 same as  in the case of $\tan\b \gg 1$.\par
The last case to consider is   of $\tan\b\simeq 1$.
Setting $\cos 2\b=0$, we get the Higgs mass matrix as
\begin{equation}
 \left (\matrix{{1\o 2}g_1+{1\o 2}g_2+\bar\xi-g & {1\o 2}(g_1-g_2) &0 \cr
{1\o 2}(g_1-g_2) &{1\o 2}g_1+{1\o 2}g_2+g+g'+h\cos2\phi& -h\sin2\phi \cr
     0 & -h\sin2\phi & g'+\bar\xi-h\cos2\phi} \right ) \ .
\end{equation}
\noindent The off diagonal components are very small compared to the
diagonal ones because $g_1\simeq g_2$ is suggested by some
 analyses[11] and $h$ is small.
Then, we get the approximate  eigenvectors  as follows:
\begin{eqnarray}
u^{(1)}&=&\{\matrix{\cos\b-\sin\b\sin\theta_{12}\cos\theta_{23}, &
             -\sin\b-\cos\b\sin\theta_{12}\cos\theta_{23},
            & \sin\theta_{12}\sin\theta_{23}}\}  ,\nonumber\\
u^{(2)}&=&\{\matrix{\sin\b\cos\theta_{23}+\cos\b\sin\theta_{12}, &
    \cos\b\cos\theta_{23}-\sin\b\sin\theta_{12}, & -\sin\theta_{23} }\} \ ,
                                                \nonumber\\
u^{(3)}&=&\{\matrix{(\sin\b\sin\theta_{23}, & \cos\b \sin\theta_{23},
              & \cos\theta_{23} }\} \ ,
\end{eqnarray}
\noindent where
\begin{eqnarray}
 \tan2\theta_{12}&=&{g_2-g_1\o \bar\xi-2g-g'-h\cos2\phi} \ , \nonumber \\
 \tan2\theta_{23}&=&{4h\sin 2\phi\o g_1+g_2+2g-2\bar\xi+4h\cos2\phi}
\simeq  {2h\sin 2\phi \o M_2^2-M_3^2}v^2 \ .
\end{eqnarray}
\noindent The Higgs scalar mass $M_1$ is  expected to be the  heaviest one and
 the $M_2$ to be the lightest one because of  $\bar\xi>3.0$
  and $g_1\sim g_2\sim g\sim |g'|\simeq O(1)$.
We estimate the effect of  $CP$ violation by considering
 the Higgs scalar with $M_2$  being  lightest one
 and then add the effect of the one with $M_3$.
 Since $\theta_{12}$ is expected to be of $O(10^{-2})$[11], we neglect terms
with $\sin\theta_{12}$ in Eq.(19).
We can calculate  the $CP$ violation parameters
${\Im}Z_i$ by fixing both values of  $h$ and $M_2/M_3$.
 We show  ${\Im}Z_1$ and ${\Im}Z_2$ in Figs.1(e) and 1(f)
 taking $M_2=200\G$, $M_3=250\G$ and $h=-0.1$.
For  both ${\Im}Z_2$ and   ${\Im}Z_1$,
 the calculating values are roughly 1/3 of the Weinberg bounds.
The relative sign between ${\Im}Z_1$ and ${\Im}Z_2$
is opposite to the one in the cases of $\tan\b\gg 1$
   and $\tan\b\ll 1$.\par
\vskip 0.5 cm
\noindent
{\bf 3.  Formulation of the neutron EDM}\par

The low energy $CP$-violating interaction is described by an effective
Lagrangian $L_{CP}$, which is generally decomposed into the local
composite operators $O_i$ of the quarks and gluons fields,
\begin{equation}
L_{CP}=\sum_i C_i(M,\mu)O_i(\mu) \ .
\end{equation}
Some authors pointed out[3,8] that the three gluon operator with the
dimension six
and the quark-gluon operator with the dimension five dominate
 EDM of the neutron in THDM. So, we study the effect of these two operators
on the neutron EDM.
 Various techniques have been developed to estimate the strong-interaction
hadronic  effects[7,8,14].
The simplest one is the NDA approach[7], but it provides at best the
  order-of-magnitude estimates.
The systematic technique has been given by Chemtob[8] for the case of the
 operator with
 the higher-dimension involving the gluon fields.
 We employ his technique to get the hadronic matrix elements of the operators.
\par
  Let us define the following  operators:
  \begin{equation}
  O_{qg}(x)=-{g_s \o 2}\bar q\sigma_{\mu\nu}\tilde G^{\mu\nu} q \ ,\qquad
  O_{3g}(x)=-{g_s^3\o 3}f^{abc}\tilde G^a_{\mu\nu}G^b_{\mu\a}G^c_{\nu\a} \ ,
\end{equation}
\noindent where $q$ denotes $u,d$ or $s$ quark.  The QCD corrected
coefficients are given by the two-loop calculations[2,3] as follows:
\begin{eqnarray}
 C_{ug}&=&-{\sqrt{2}G_F m_u(\mu)\o 128\p^4}g_s^2(\mu)[f(z_t)+g(z_t)]\Im Z_2
      \left ({g_s(\mu)\o g_s(M)}\right )^{-{74\o 23}} \ , \nonumber \\
 C_{dg}&=&-{\sqrt{2}G_F m_{d}(\mu)\o 128\p^4}g_s^2(\mu)
[f(z_t)\tan^2\b \Im Z_2 -g(z_t)\cot^2\b \Im Z_1]
      \left ({g_s(\mu)\o g_s(M)}\right )^{-{74\o 23}} \ ,  \nonumber\\
 C_{3g}&=&{\sqrt{2}G_F \o 256\p^4}h(z_t)\Im Z_2
      \left ({g_s(\mu)\o g_s(M)}\right )^{-{108\o 23}} \ ,
\end{eqnarray}
\noindent where $z_t=(m_t/m_H)^2$
and we omitt the upper-indices $(k)$ defined in Eq.(2).
The function $f(z_t)$, $g(z_t)$
and $h(z_t)$ are the two-loop integral function,
which are defined in Refs.[4,5,15].  The  $C_{sg}$ coefficient is
same as $C_{dg}$ except for the quark mass.
In our practical calculation, the modification to account for the passage
through the $b$ and $c$ quarks thresholds involves the replacement
\begin{equation}
      \left ({g_s(\mu)\o g_s(M)}\right )^{{n\o 23}} \longrightarrow
       \left ({g_s(m_b)\o g_s(M)}\right )^{{n\o 23}}
       \left ({g_s(m_c)\o g_s(m_b)}\right )^{{n\o 25}}
        \left ({g_s(\mu)\o g_s(m_c)}\right )^{{n\o 27}} \ .
\end{equation}
\par
The hadronic matrix elements of the two operators
 are approximated by the intermediate states with the single nucleon pole
 and the  nucleon plus one pion. Then, the nucleon matrix elements are
 defined as
\begin{eqnarray}
  \langle N(P)|O_i(0)|N(P)\rangle &=& A_i\bar U(P)i\r_5 U(P) \ , \nonumber\\
  \langle N(P')|O_i|N(P)\p(k)\rangle &=& B_i\bar U(P')\tau^a U(P) \ ,
\end{eqnarray}
\noindent
where $U(P)$ is the normalized nucleon Dirac spinors
               with the four momuntum $P$.
Using $A_i$ and $B_i(i=ug,dg,sg,3g)$, the neutron EDM, $d_n^\r$,
are written as
\begin{equation}
  d_n^\r={e\mu_n\o 2 m_n^2}\sum_i C_i A_i +
  F(g_{\p NN},m_n,m_\p )\sum_i C_i B_i    \ ,
\end{equation}
\noindent where $\mu_n$ is the neutron anomalous magnetic moment.
 The function $F(g_{\p NN},m_n,m_\p )$ was derived by calculating the pion
and nucleon loop corrections using the chiral Lagrangian
for the coupled $N\p\r$ and in given in Appendix A of Ref.[8].
Here, the dimensional regularization with the  standard $\bar {MS}$ scheme
is  used for defining the finite parts of the divergent integrals.
The coefficients $A_i$ and $B_i$ were given by the use of the large $N_c$
current algebra and the $\eta_0$ meson dominance[8].
Then,  we have
\begin{equation}
 A_i=f_i g_{\eta_0 NN} \ ,\hskip 1.5 cm B_i=-{4(m_u+m_d)a_1f_i\o F_\p F_0} \ ,
\end{equation}
\noindent with $a_1=-(m_{\Sigma^0}-m_{\Sigma})/(2m_s-m_u-m_d)\simeq -0.28$
and $F_\p=\sqrt{2/3}F_0=0.186\G$,
 where  $f_i$ is defined as
\begin{equation}
 \langle \eta_0(q)| O_i(0)|0 \rangle \equiv f_i q^2 \ .
\end{equation}
\noindent
 The values of  $f_i$ were derived by using QCD sum rules as follows[8]:
\begin{equation}
 f_{qg}=-0.346\G^2 \ , \hskip 2 cm f_{3g}=-0.842 \G^3\ ,
\end{equation}
\noindent
 where $f_{qg}$ denotes the flavor singlet coupling.
\par
Now, we can calculate the neutron EDM. Our inputs parameters
are
\begin{eqnarray}
& &\Lambda_{QCD}=0.26\G \ ,  \qquad (m_u,m_d,m_s)=(5.6,9.9, 200) {\rm MeV}\ ,
   \qquad    \mu=m_n \ ,    \nonumber \\
& & M=m_t=150 \G \ , \qquad  g_{\p NN}=13.5 \ , \qquad g_{\eta_0 NN}=0.892 \ .
\end{eqnarray}
 \noindent  Here, it is useful to comment on the value of $\mu$.
 As the smaller $\mu$ is taken, the QCD suppression factor increases, and then,
 the predicted neutron EDM decreases.
Although we do not have the reliable principle to fix $\mu$ in the leading-log
approximation of QCD, we tentatively
   take $\mu=m_n$, which leads $\a_s(\mu)=0.54$. If we take  $\mu=0.6\G$,
             which gives rather large $\a_s(\mu)=0.83$,
as used by Chemtob[8], our predicted neutron EDM
will be reduced by a factor $2\sim 4$.
\par
\vskip 0.5 cm
\noindent
{\bf 4.  Numerical analyses of the neutron EDM}\par

 We show the numerical results in this section.
Since the $CP$ violation parameters $\Im Z_i$ have been estimated
in the three cases of $\tan\b$, the neutron EDM is predicted
 for each case of $\tan\b$.
 Since our results are proportional to $\sin 2\phi$, we take the maximal
 case $\phi=\pi/4$ in showing the numerical results. We show
the contribution of the four operators $O_{ug}$,  $O_{dg}+O_{sg}$ and $O_{3g}$
 on the neutron EDM, respectively.
At first, we show the predicted neutron EDM
in the region of $5\leq\tan\b\leq 10$,
which corresponds to the case of $\tan\b\gg 1$, in  Fig.2(a), where
the combined experimental upper bound of the neutron EDM[16],
$8\times 10^{-26} {\rm e\cdot cm}$, is shown by the horizontal dotted line.
  The two lightest Higgs scalars have been
                 taken into account in our calculations.
 Defining the two lightest Higgs scalar masses  to be $m_{H1}$ and $m_{H2}$,
  we fixed tentatively  $m_{H1}=200\G$ and $m_{H2}=250\G$,
   which correspond to $h=-0.37$.
In Fig.2(a), the contributions of $O_{ug}$ and $O_{3g}$ are
shown  multiplying them by the factor $100$ because they are very small.
  It is noted that the signs of these two contributions are opposite,
and  they almost cancel each other. The main contribution
follows from the one of $O_{dg}+O_{sg}$, in which the operator $O_{sg}$
is dominant due to the $s$-quark mass. This contribution is constant versus
 $\tan\b$ since the $\tan\b$ dependence  of $C_{dg}+C_{sg}$
 disappears as seen in the Eqs.(2) and (23), and then overlapps perfectly
to the total EDM(solid line) in Fig.2(a).
Thus,  the $O_{sg}$ operator dominates the neutron EDM in the case of
$\tan\b\gg 1$.
\begin{center}
 \unitlength=0.7 cm
 \begin{picture}(2.5,2.5)
  \thicklines
  \put(0,0){\framebox(3,1){\bf Fig.2(a)}}
 \end{picture}
\end{center}
\par
  As the mass difference of these
two Higgs scalar masses becomes smaller, the neutron EDM  is considerably
reduced since the second Higgs scalar exchange  contributes
in the opposite sign to the  lightest Higgs scalar one.
In Fig.2(b), we show the the predicted neutron EDM versus $m_{H1}/m_{H2}$
in the case of $\tan\b=10$ with $m_{H1}=200$ and $400\G$.
As far as  $m_{H1}/m_{H2}\geq 0.7$($|h|\leq 0.68$),
 the predicted value lies under the
experimental upper bound. Thus, it is found that the second lightest
Higgs scalar also significantly contributes to  $CP$ violation.
\begin{center}
 \unitlength=0.7 cm
 \begin{picture}(2.5,2.5)
  \thicklines
  \put(0,0){\framebox(3,1){\bf Fig.2(b)}}
 \end{picture}
\end{center}
\par
The neutron EDM in the case of  $\tan\b\ll 1$ is shown in Fig.3(a), where
we take the region of $\tan\b\leq 0.25$.
The contributions of $O_{ug}$ and $O_{3g}$ become very large due to
 the large $\Im Z_2$.
 However, these contribute to the neutron EDM  in opposite signs,
so they almost cancel each other in the region of $1\gg \tan\b\geq 0.1$
 as shown in Fig.3(a).
The remaining contribution is the one of $O_{dg}+O_{sg}$, which is
 constant versus $\tan\b$.  In the region of $\tan\b\leq 0.1$,
 the  cancelation between $O_{ug}$ and $O_{3g}$ is violated and
 the contribution of  $O_{ug}$ donimates the neutron EDM in the region of
$\tan\b\ll 0.1$.
\begin{center}
 \unitlength=0.7 cm
 \begin{picture}(2.5,2.5)
  \thicklines
  \put(0,0){\framebox(3,1){\bf Fig.3(a)}}
 \end{picture}
\end{center}
\par
In Fig.3(b), we show the  predicted neutron EDM versus $m_{H1}/m_{H2}$
in the case of $\tan\b=0.1$.
The allowed parameter region of $m_{H1}/m_{H2}$ is obtained by the experiment
and is   $m_{H1}/m_{H2}\geq 0.95$($|h|\leq 0.07$).
In othe words,
 the second lightest Higgs scalar mass should be close to the lightest one.
We want to note that the predicted EDM with $m_{H1}=400\G$
is larger than the one with $m_{H1}=200\G$
         in the region of $m_{H1}/m_{H2}\geq 0.5$.
 The  cancelation between $O_{ug}$ and $O_{3g}$ is violated and
 the contribution of  $O_{3g}$ dominates the neutron EDM in the case of
$m_{H1}=400\G$ at $\tan\b=0.1$.
 Thus, one should carefully analyze the signs and magnitudes of
the contribution of
$O_{ug}$,  $O_{dg}+O_{sg}$ and $O_{3g}$ operators in the case of
           $\tan\b\ll 1$ since those  sensitively depend on the
 values of $m_{H1}$, $m_{H2}$ and $\tan\b$.
\begin{center}
 \unitlength=0.7 cm
 \begin{picture}(2.5,2.5)
  \thicklines
  \put(0,0){\framebox(3,1){\bf Fig.3(b)}}
 \end{picture}
\end{center}
\par
The neutron EDM in the case of $\tan\b\simeq 1$ is shown
in Fig.4(a).
Since the parameter $h$ is independent of
 the Higgs scalar mass difference in contrast to the
above two cases,  we fix $h=-0.1$ as a typical value with
$m_{H1}=200\G$ and $m_{H2}=250\G$.
The contributions of $O_{ug}$ and $O_{3g}$ are
shown  multiplying them by the factor $10$.
Similarly to be former cases, the signs of these two contributions
are opposite
and  cancel each other, and so the dominant contribution
 is the one of $O_{dg}+O_{sg}$,  which overlapps perfectly
to the total EDM(solid line) in Fig.4(a).
\begin{center}
 \unitlength=0.7 cm
 \begin{picture}(2.5,2.5)
  \thicklines
  \put(0,0){\framebox(3,1){\bf Fig.4(a)}}
 \end{picture}
\end{center}
\par
In Fig.4(b),  the predicted neutron EDM is shown versus $m_{H1}/m_{H2}$
in the case of $\tan\b=1$ with $h=-0.05, -0.1$.
In the region of $m_{H1}/m_{H2}=0.5\sim 0.9$,
the predicted value is over the experimental upper bound in the case of
$m_{H1}=200\G$ with $h=-0.1$. Thus,
the magnitude of  $|h|$  is rigorously restricted by the experimental
upper bound of the neutron EDM.
In both regions of the large and small $m_{H1}/m_{H2}$,
the predicted neutron EDM is reduced.
At  $m_{H1}/m_{H2}\simeq 1$, the cancellation mechanism by the
second lightest Higgs scalar operates well, while
  around $m_{H1}/m_{H2}\simeq 0$,
the large mass difference of the two Higgs scalars leads to the small
$\theta_{23}$ as seen in Eq.(20).
\begin{center}
 \unitlength=0.7 cm
 \begin{picture}(2.5,2.5)
  \thicklines
  \put(0,0){\framebox(3,1){\bf Fig.4(b)}}
 \end{picture}
\end{center}
\par
In all cases of $\tan\b$, the contribution of $O_{dg}+O_{sg}$
dominate the neutron EDM. The effects of $O_{ug}$ and $O_{3g}$
seem to become large only in the region of $\tan\b\ll 1$ although
these cancel each other considerably.\par
\vskip 0.5 cm
\noindent
{\bf 5.  Conclusion}\par

We have studied  the effects  of the
 four operators $O_{ug}$,  $O_{dg}+O_{sg}$ and $O_{3g}$ on the neutron EDM.
The contribution of $O_{sg}$  dominates over that of other operators
except for the region of $\tan\b\ll 1$. Moreover,
the contributions of $O_{ug}$ and $O_{3g}$ cancel out each other
due their opposite signs.  This qualitative situation does not
depend on the detail of the strong interaction hadronic model.
Actually, in the NDA approximation[7] of the hadronic effect,
 the effects of the two operators almost cancel out
    although the predicted EDM is smaller than ours by a factor $2\sim 3$.
Thus, the Weinberg's three gluon operator
is not a main source of the neutron EDM in THDM.
Of course, Weinberg's  operator may be dominant one in the other models
beyond SM,  which
we will investigate elsewhere.
The CP violation mainly follows from the two light neutral Higgs  scalar
exchanges. Since these two exchange contributions are of
opposite signs, the $CP$ violation is considerably reduced
 if the mass difference of the two Higgs scalars
        is within the order of $O(50\G)$.\par
 Since our results have been shown by taking  $\sin\phi=\p/4$,
 for an arbitrary $\phi$ our predicted neutron EDM
is simply scaled by the factor $\sin2\phi$.
  This factor is expected
to be of the order one unless $\phi$ is suppressed by an unknown mechanism
        in THDM.
 Therefore, our results remain unchanged qualitatively.\par
 Since our predicted neutron EDM lies around the present
 experimental bound,
  its experimental improvement may reveal the new physics
beyond SM.
\par
\vskip 0.5 cm
\noindent
\centerline{\bf Acknowledgments}\par
This research is supported  by the Grant-in-Aid for Scientific Research,
Ministry of Education, Science and Culture, Japan(No.05228102).
\newpage
\centerline{\large \bf References}
\noindent
[1] M.Kobayashi and T.Maskawa, Prog. Theor. Phys. {\bf 49}(1973) 652.\par
\noindent
[2] S. Weinberg, Phys. Rev. Lett. {\bf 63}(1989)2333.\par
\noindent
[3] J.F. Gunion and D. Wyler, Phys. Letts. {\bf 248B}(1990)170.\par
\noindent
[4] A. De R$\acute u$jula, M.B. Gavela, O. P$\grave e$ne and F.J. Vegas,
    Phys. Lett. {\bf 245B}(1990)640; \par
    N-P. Chang and D-X. Li, Phys. Rev. {\bf D42}(1990)871;\par
  D.Chang, T.W.Kephart, W-Y.Keung and T.C.Yuan,
                    Phys.Rev.Lett. {\bf 68}(1992)439;\par
 M. J. Booth and G. Jungman, Phys. Rev. {\bf D47}(1993)R4828.\par
\noindent
[5] S.M. Barr and A. Zee, Phys. Rev. Lett. {\bf 65}(1990)21;\par
   S.M. Barr, Phys. Rev. Lett. {\bf 68}(1992)1822;
            Phys. Rev. {\bf D47}(1993)2025.\par
\noindent
[6] For a text of Higgs physics see J.F. Gunion, H.E. Haber, G.L.Kane
    and S. Dawson,\par
  {\it "Higgs Hunter's Guide"},  Addison-Wesley, Reading, MA(1989).
\par
\noindent
[7] A. Manohar and H. Georgi, Nucl. Phys. {\bf B234}(1984)189.\par
\noindent
[8] M. Chemtob, Phys. Rev. {\bf D45}(1992)1649.\par
\noindent
[9] S. Weinberg, Phys. Rev. {\bf D42}(1990)860.\par
\noindent
[10]M. Chemtob, Z. Phys. {\bf C60}(1993)443.\par
\noindent
[11]M.A. Luty, Phys. Rev. {\bf D41}(1990)2893;\par
 C.D. Froggatt, I.G. Knowles, R.G. Moorhouse, Phys. Lett. {\bf 249B}(1990)273;
   \par  Nucl. Phys. {\bf B386}(1992)63.\par
\noindent
[12]T.Hayashi, M.Matsuda and M.Tanimoto,
          Prog. Theor. Phys. {\bf 89}(1993)1047;\par
   T.Hayashi, M.Matsuda and M.Tanimoto, preprint AUE-02-93(1993).\par
\noindent
[13]E. Thorndike(CLEO Collabo.), Talk given at the Meeting of the American
    \par Physical Society(Washington D.C., 1993);\par
  R. Ammar et al., CLEO Collaboration, Phys. Rev. Lett. {\bf 71}(1993)674.\par
\noindent
[14]X-G. He, B.H.J. Mckellar and S. Pakvasa, Mod. Phys. {\bf A4}(1989)5011;\par
  I.I. Bigi and N.G. Uraltsev, Nucl. Phys. {\bf B353}(1991)321.\par
\noindent
[15]D.A. Dicus, Phys. Rev. {\bf D41}(1990)999;\par
  D. Chang, W-Y. Keung and T.C. Yuan, Phys. Lett. {\bf 251B}(1990)608.\par
\noindent
[16] Particle Data Group, Phys. Rev. {\bf D45}(1992) II-1.\par

\newpage
\centerline{\large \bf Figure Captions}\par
\vskip 0.5 cm
\noindent
{\bf Fig.1}: The predicted $CP$ violation factors in the case of
 $\phi=\p/4$.  The solid curves show
 (a)$\Im Z_1$ and (b)$\Im Z_2$ in  $\tan\b=5\sim 10$,
 (c)$\Im Z_1$ and (d)$\Im Z_2$ in  $\tan\b=0\sim 0.3$,
   (e)$\Im Z_1$ and (f)$\Im Z_2$ in  $\tan\b=0.8\sim 1.2$
with $M_2=200\G$, $M_3=250\G$ and $h=-0.1$.
The dashed curves denote the upper bounds given by Weinberg.\par
\vskip 0.3 cm
\noindent
{\bf Fig.2(a)}: The predicted neutron EDM in $\tan\b=5\sim 10$
with $m_{H1}=200\G$ and $m_{H2}=250\G$.
The dotted curve and dashed curve denote the contribution by
 $O_{ug}$ and  $O_{3g}$, respectively. The contribution of
$O_{dg}+O_{sg}$ overlapps the total neutron EDM shown by the solid line.
The horizontal dotted line denotes the experimental upper bound.
\par
\vskip 0.3 cm
\noindent
{\bf Fig.2(b)}: The $m_{H1}/m_{H2}$ dependence of the neutron EDM in
 $\tan\b=10$  with $m_{H1}=200,400\G$.
\par
\vskip 0.3 cm
\noindent
{\bf Fig.3(a)}: The predicted neutron EDM in  $\tan\b=0\sim 0.3$
with $m_{H1}=200\G$ and $m_{H2}=250\G$.
The notations are same as in Fig.2(a). The dashed horizontal line
 denotes the contribution by $O_{dg}+O_{sg}$.
\par
\vskip 0.3 cm
\noindent
{\bf Fig.3(b)}: The $m_{H1}/m_{H2}$ dependence of
 the neutron EDM in
 $\tan\b=0.1$  with $m_{H1}=200,400\G$.\par
\vskip 0.3 cm
\noindent
{\bf Fig.4(a)}: The predicted neutron EDM in  $\tan\b=0.8\sim 1.2$
with $m_{H1}=200\G$ and $m_{H2}=250\G$.
       The notations are same as in Fig.2(a).
\par
\vskip 0.3 cm
\noindent
{\bf Fig.4(b)}: The $m_{H1}/m_{H2}$ dependence of the neutron EDM
in $\tan\b=1$ with $m_{H1}=200, 400\G$ and $h=-0.05, -0.1$.\par
\end{document}